# Strategies For Non-Planar Configurations Of Geostationary Tethered Collecting Solar Power Satellite Systems

Francisco J. T. Salazar[1,*] and Antonio B. A. Prado[1]

[1]Orbital Mechanics Department, National Institute for Space Research, Av. dos Astronautas 1.758, São José dos Campos, SP 12227-010, Brazil
[*]e7940@hotmail.com

**ABSTRACT**

To collect additional solar energy during the hours of darkness and to overcome the limited Terrestrial solar power due to the diurnal day-night cycle, the concept of a Geostationary Tethered Collecting Solar Power Satellite System has been proposed by several authors in the last years. This tethered system consists of a long tether used to link two bodies: a single large panel with a capability of collecting solar energy, and an Earth-pointing microwave transmitting satellite. In this manner, the solar energy would be collected directly from the space and beamed back down to any point on Earth. Planar configurations, when the panel and the microwave transmitting satellite are placed on geostationary orbits, have been usually investigated to maintain the tethered system around the Earth. However, this configuration implies that the panel and the microwave transmitting satellite must to orbit the Earth in exactly the same orbital plane of all geostationary satellites. However, the geostationary band are getting more crowded every year. Thus, non-planar configurations in geostationary orbits (e.g. Displaced Geostationary Orbits, Light-Levitated Geostationary Cylindrical Orbits) have been studied by various authors to increase the number of slots over a particular longitude. In this paper, we consider both in-plane and out-of-plane librations, and derive a linear control strategy for a non-planar configuration of a tethered collecting solar power system. The microwave transmitting satellite is placed on a geostationary orbit, and the panel, perturbed by the solar radiation pressure, achieves a specific stable out-of-plane configuration applying a tension control law in conjunction with out-of-plane thrusting. The tension control law and the out-of-plane thrust are derived from the linearization of the governing nonlinear differential equations about an out-of-plane equilibrium point, so that the equilibrium solution is asymptotically stable. The numerical simulations show that the linear control strategy designed from this approach performs well for the nonlinear system, and the solar power system can be guided to the desired non-planar configuration.

## Introduction

The concept of a Tethered Satellite System (TSS) consists of a long tether used to link two or more satellites as they orbit a central body (e.g. the Earth). In later decades, many different space applications have been proposed, such as micro and variable-$g$ experiments, orbit transfer, spacecraft formation, energy transmission, space debris removal, and many others[1]. However, one of the primary issues in space tether missions is the fast deployment/retrieval of attached payloads. Controlling tether deployment/retrieval is extremely complicated due to the Coriolis force makes the deployment/retrieval operations to be unstable. Actually, the retrieval operation is inherently unstable[2–4]. Additionally, the nonlinear dynamics of the problem can generate quasi-periodic and chaotic motions in a two-body TSS[4–17]. Many researchers looked at this problem, investigating the dynamics and control of a TSS during deployment, station keeping, and retrieval stages. Various methods for controlling tether deployment/retrieval have been proposed, such as tension control laws, length rate control algorithms, Liapunov's second method, optimal control strategies, elastic tethers, and out-of-plane thrusting[4,17–39].

Because of the complexity of the problem, a TSS consisted of two spacecraft (a base-satellite and a sub-satellite) linked by a rigid but massless tether, that is revolving around the Earth in a Keplerian circular orbit, is considered in this study. Herein, we analyze the efficiency of a nonlinear tension control law designed from a Lyapunov function, and applied to the problem of tether deployment and retrieval when the initial state emerges from the three-dimensional quasi-periodic and chaotic zones of the TSS dynamics. Lyapunov's second method has been used to select a stable tension control law during deployment and retrieval in the three-dimensional case, such that the quasi-periodic and chaotic motions are suppressed and the system is steered close to the local vertical direction. Although innumerable literature have been published about the deployment and retrieval dynamics of tether connected two-body systems, the control of the chaotic behavior observed in the coupled pitch-roll motion has not been sufficiently explored. In this sense, the simulations performed in this study, show that the selected nonlinear tension control law performs well, but terminal oscillations of the roll motion are encountered during retrieval due to this process is inherently unstable, as mentioned above.

## Results

In this study, we consider the deployment and retrieval of a TSS consisting of two point size bodies of masses $M$ (base-satellite) and $m$ (sub-satellite) connected by a rigid and massless tether. The base-satellite is traveling in a Keplerian circular orbit around the Earth, at an altitude of 220 km, and an orbital angular velocity $\omega = 1.1804 \times 10^{-3}$ rad/s. The maximum tether length is 1 km. The masses of the satellites are set as $M = 1,000$ kg and $m = 50$ kg. The initial conditions for the deployment of the sub-satellite are $(\rho_0, \rho'_0, \phi_0, \phi'_0, \theta_0, \theta'_0) = \left(0.01, 0.5, \frac{\pi}{2}, 0, 0, C\right)$, where $\rho$ denotes the instantaneous non-dimensional length of the tether, and $\phi$-$\theta$ denote the pitch-roll angles, respectively. Similarly, $\rho'$, $\phi'$, $\theta'$ denote the non-dimensional velocity of the variables $\rho$, $\phi$, $\theta$, respectively. The initial conditions $\rho_0, \rho'_0, \phi_0$, and $\phi'_0$ are the same ones used in other studies[25,29,38]. On the other hand, the initial roll motion $\theta'_0$ was set such that quasi-periodic and chaotic solutions appear when the tether length is constant.

The deployment and retrieval that take into consideration out-of-plane motion are more complicate, and further studies and experiments have been carried out by many researchers, that investigate the three-dimensional dynamics and control of TSS. In this connection, innumerable control strategies have been proposed for the in-plane deployment and retrieval[19,20,23–26,28,29,31,38]. Among them, the tension feedback control laws, as well as the length rate schemes, have been proven to be the most effective due to their simplicity and practicality[40].

In this sense, a nonlinear feedback control law was designed using a Liapunov function (named mission function), which showed excellent controlled response of the tethered sub-satellite system. The Lyapunov function was defined to be positive-definite and to be zero when the deployment and retrieval are essentially completed. A non-dimensional tension control $u$ was then selected to reduce the value of the mission function as time increases, i.e. the time derivative of the mission function was forced by the control process to be negative-definite during deployment and retrieval phases. In this manner, a stable nonlinear control law could be designed from a positive-definite mission function, which was applied equally to accomplish both the deployment and the retrieval cases.

The desired final conditions are $\left(\rho_f, \rho'_f, \phi_f, \phi'_f, \theta_f, \theta'_f\right) = \left(\rho^*_f, 0, \frac{\pi}{2}, 0, 0, 0\right)$, where $\rho^*_f = 1.0$ for deployment and $\rho^*_f = 0.01$ for retrieval. In this manner, the effect of variable length of the tether on the three-dimensional solutions is evaluated using the nonlinear tension control, which must satisfy the necessary positiveness condition, i.e. $u \geq 0$, at any time to confirm the workability of the control law.

Figures 1 to 3 show the effect of the variable length of the tether, respectively. Figures 1 to 3 show the time history of the pitch and roll angles. Similarly, Figure 1 to 3 show the time history of the non-dimensional tether length and tension. It is seen that the non-dimensional tension control $u$ is always positive in both cases, and the deployment and retrieval phases are essentially complete in about five orbits. There exists a station-keeping phase when the deployment and retrieval phases are complete, as shown in Figs. 1 to 3. In the station-keeping phase, the tension control tends to stabilize about $u = 3$ ($\tau = 0.21$ N) when $\rho \pm 1$, and $u = 0$ when $\rho \theta.01$. Although the tension control performs well, and the tether achieves the desired final lengths, the pitch and roll angles oscillate about the equilibrium vertical position during the transfer over from station-keeping to retrieval, as shown in Figs. 1 to 3. However, as soon as the tension control causes the tether to retrieve from $\rho = 1$ to $\rho = 0.01$, we have that the length rate settles down to $\rho' = 0$, the pitch motion returns to the equilibrium, and the roll motion settles down to a sinusoidal oscillation of constant amplitude, corresponding to a limit cycle[4]. In this connection, an out-of-plane thrusting for the retrieval phase have been proposed for several authors to help stabilize this process[21,23,30,41].

## Discussion

In the station-keeping phase of a TSS, the coupled pitch-roll motion is highly nonlinear and possibly chaotic. Actually, the local vertical position, which is a common initial state for fast deployment/retrieval of payloads in a TSS, can change from a stable equilibrium position to quasi-periodic or chaotic trajectory if roll motion is initially excited. As a main contribution of the present research, a Lyapunov approach was used in this paper to determine a new nonlinear tension control law for controlling tether deployment and retrieval. In this method, the tension in the tether controls the length of the tether.

Considering the pitch and roll motions, the tethered sub-satellite was deployed from a perturbed vertical position, such that quasi-periodicity and chaos occur when the length of the tether is constant. The tension control was designed to extend or retrieve the tether to the required maximum or minimum lengths, respectively, while suppressing the quasi-periodic and chaotic oscillations of the tether. From the Lyapunov stability analysis performed about the desired final position, it was shown that the final vertical position is stable, but not asymptotically stable. Although a fast deployment/retrieval was accomplished by the nonlinear tension control law outlined in this study, and the tension was capable of controlling the three-dimensional motion, and to guide the system to the desired final state, terminal small oscillations of the roll motion can be seen during the retrieval phase. The result is a sinusoidal oscillation of constant amplitude about the local vertical. New control laws will be designed to hold the pitch and roll angles constant during the terminal phase of retrieval by using more sophisticated models of TSS.

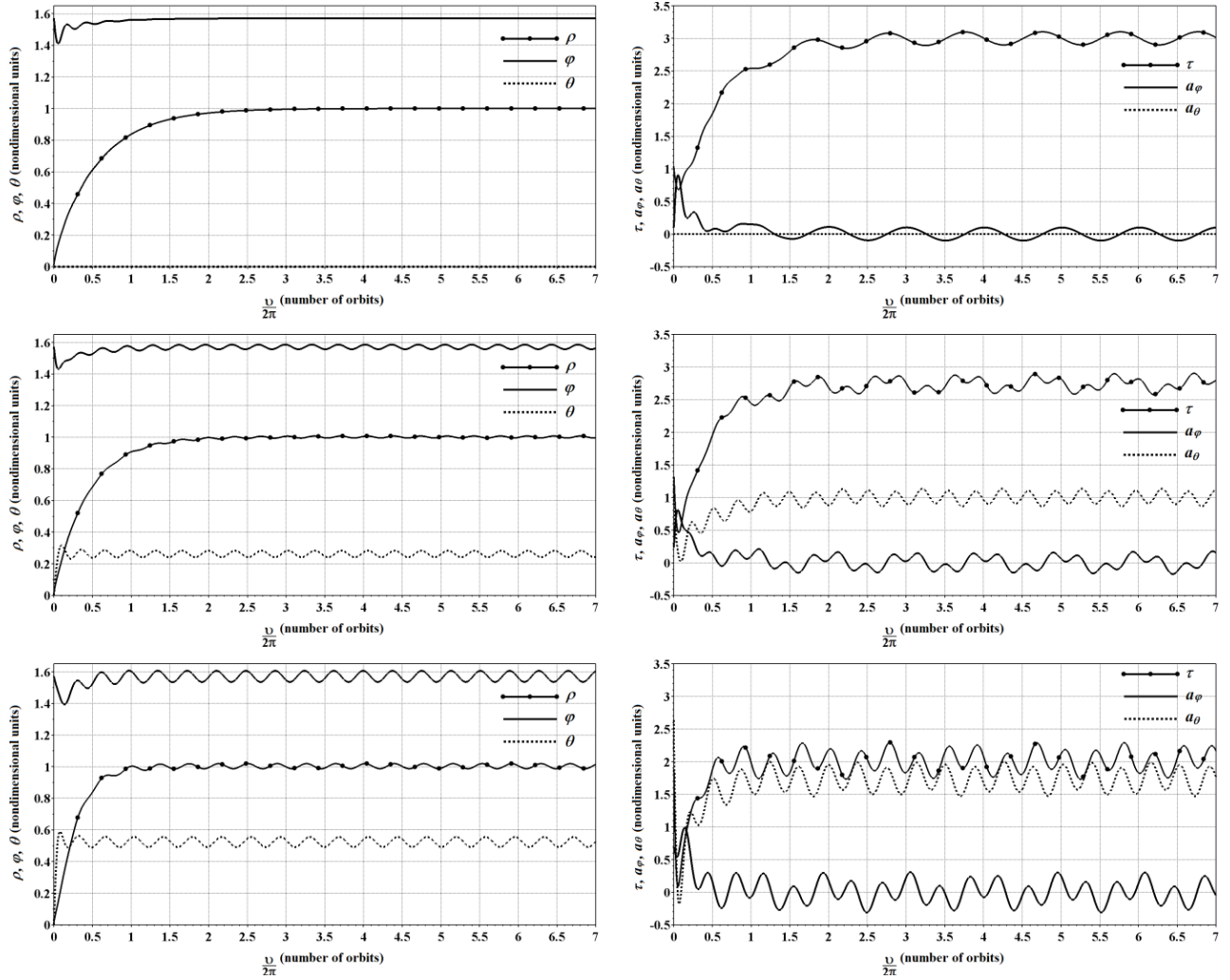

**Figure 1.** Controlled Tether Deployment and Retrieval.

## Methods

Assuming that the base-satellite follows a Keplerian circular orbit with true anomaly $\upsilon$ and orbital angular velocity $\omega$, and the sub-satellite's mass $m$ is far smaller than the base-satellite's mass $M$ (i.e., $m \ll M$), then the relative motion of the sub-satellite with respect to the base-satellite is described by the following equations[31]:

$$\ddot{\ell} - \ell\dot{\theta}^2 - \ell(\omega+\dot{\phi})^2\cos^2\theta - \ell\omega^2\left[3\sin^2\phi\cos^2\theta - 1\right] = -\frac{T}{m}, \tag{1}$$

$$\ddot{\phi} + 2\frac{\dot{\ell}}{\ell}(\omega+\dot{\phi}) - 2(\omega+\dot{\phi})\dot{\theta}\tan\theta - 3\omega^2\sin\phi\cos\phi = 0, \tag{2}$$

$$\ddot{\theta} + 2\frac{\dot{\ell}}{\ell}\dot{\theta} + (\omega+\dot{\phi})^2\cos\theta\sin\theta + 3\omega^2\sin^2\phi\sin\theta\cos\theta = 0, \tag{3}$$

where $\ell$ is the instantaneous length of the rigid massless tether, and $\phi$-$\theta$ denote the pitch-roll angles, respectively. $T$ is the magnitude of the tension tether force, and overdot denotes differentiation with respect to time $t$. Note that the damping term in equations (2) and (3) is proportional to $\dot{\ell}$. Thus, the pitch and roll motions are positively damped and stable during deployment, but they are negatively damped and unstable during retrieval, as shown in Figs. 1 to 3. On the other hand, these equations can

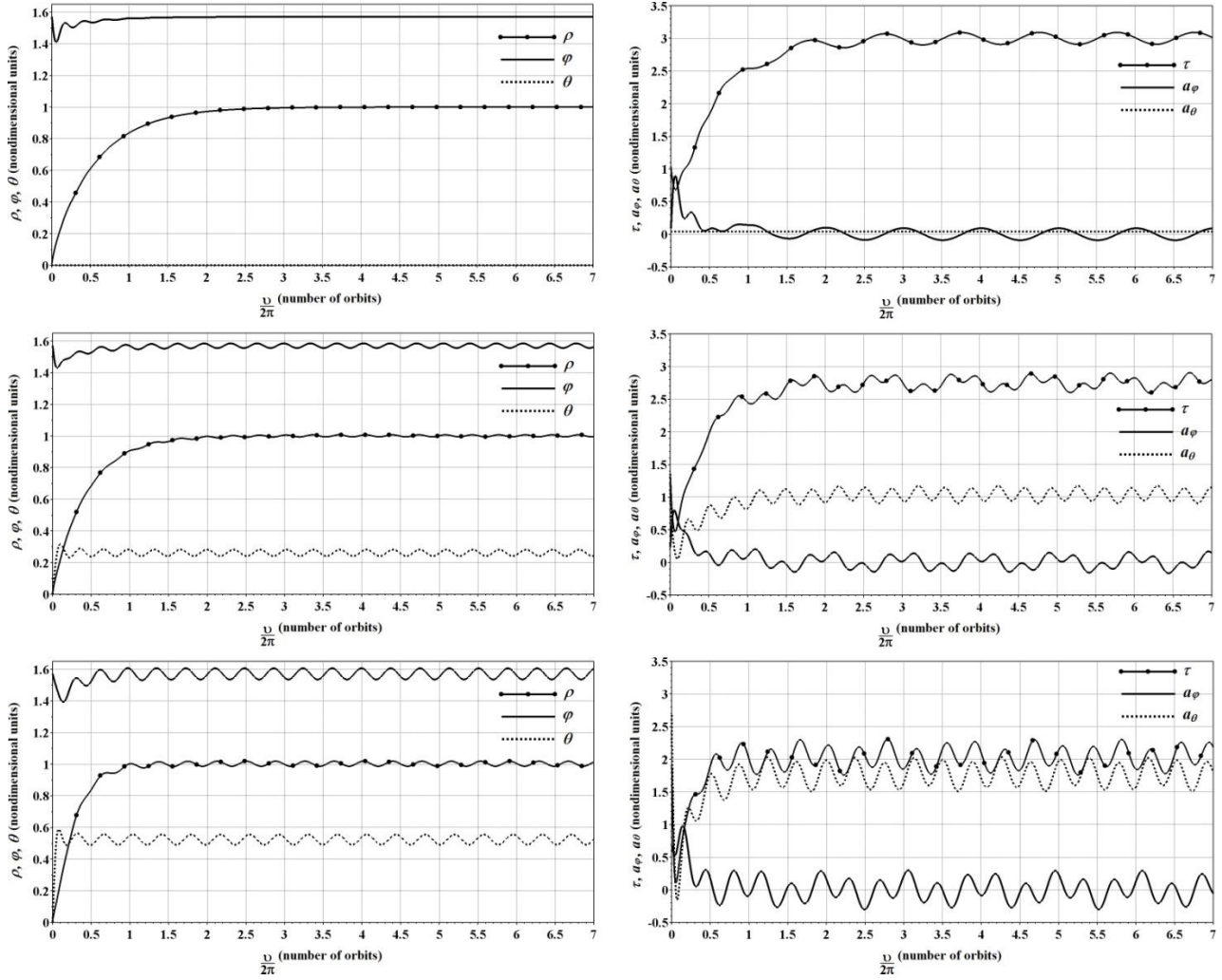

**Figure 2.** Controlled Tether Deployment and Retrieval.

be non-dimensionalized by using the following non-dimensional variables:

$$\upsilon = \omega t, \quad \rho = \frac{\ell}{L}, \quad u = \frac{T}{m\omega^2 L},$$

where $L$ is the maximum tether length. The non-dimensional equations are given by the following:

$$\rho'' - \rho\theta'^2 - \rho\left[\left(1+\phi'\right)^2 \cos^2\theta - \rho\left(3\sin^2\phi\cos^2\theta - 1\right)\right] = -u, \quad (4)$$

$$\phi'' + 2\frac{\rho'}{\rho}\left(1+\phi'\right) - 2\left(1+\phi'\right)\theta'\tan\theta - 3\sin\phi\cos\phi = 0, \quad (5)$$

$$\theta'' + 2\frac{\rho'}{\rho}\theta' + \left[\left(1+\phi'\right)^2 \cos\theta\sin\theta + 3\sin^2\phi\sin\theta\cos\theta\right] = 0, \quad (6)$$

where prime refers to differentiation with respect to the true anomaly $\upsilon$. When $\rho' = 0$ (station keeping phase) the system has three equilibrium points: the local vertical with $\phi = \pi/2$, $\theta = 0$, the local horizontal with $\phi = 0$, $\theta = 0$, and the orbit normal with $\theta = \pi/2$. Only the local vertical position is stable[8,9]. Additionally, there exists a non-dimensional integral $C$ (i.e. $C' = 0$) for the coupled system in the station keeping phase ($\rho' = 0$), given by[8]

$$C = \theta'^2 + \cos^2\theta\left(\phi'^2 - 1 - 3\sin^2\phi\right) + 4, \quad (7)$$

where $0 \leq C \leq 4$, with $C = 0$ and $C = 4$ at the local vertical and orbit normal equilibrium points, respectively.

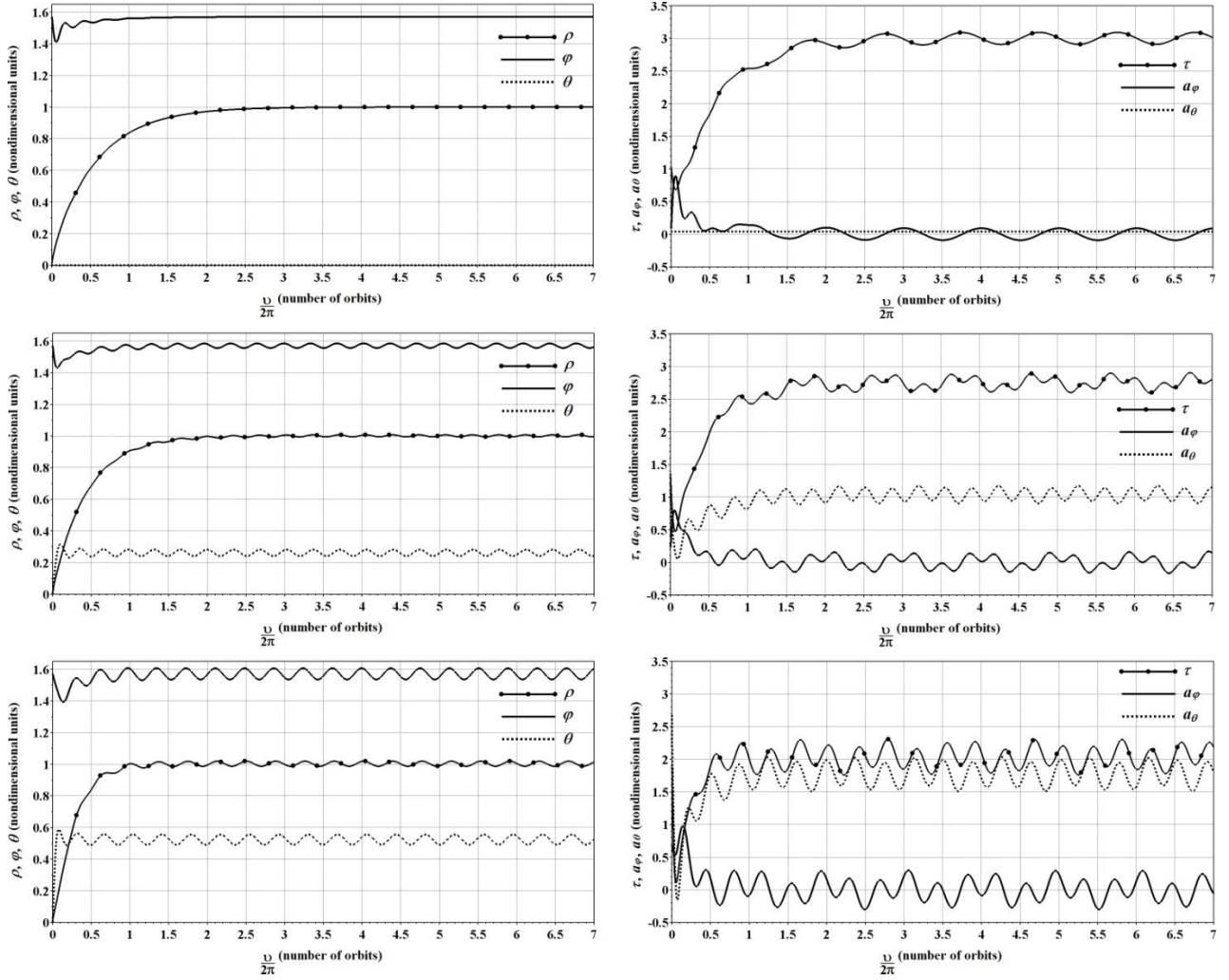

**Figure 3.** Controlled Tether Deployment and Retrieval.

The following positive definite function, which is related to the integral $C$, is selected as the trial Lyapunov function:

$$V = \frac{1}{2}\left[\rho'^2 + K_1\left(\rho - \rho_f\right)^2 + 3\rho^2 C^2\right], \qquad (8)$$

where $\rho_f$ is the desired final value of $\rho$, $K_1$ is a positive constant (non-dimensional gain), $V \geq 0$, and $V = 0$ if and only if at $\rho = \rho_f$, $\phi = \pm(2n+1)\frac{\pi}{2}$, $\theta = \pm n\pi$, $\rho' = 0$ $\phi' = 0$, $\theta' = 0$. The time derivative of $V$ is given by

$$V' = \rho'\left[\rho'' + K_1\left(\rho - \rho_f\right) + 3\rho C\left(C - 4\left(\theta'^2 + \phi'\cos^2\theta\left(1+\phi'\right)\right)\right)\right]. \qquad (9)$$

Substituting equation (4) in (9), we obtain

$$u = u_1 + u_2 + u_3, \qquad (10)$$

where

$$u_1 = K_1\left(\rho - \rho_f\right) + K_2\rho', \qquad (11)$$

$$u_2 = \rho\left[\theta'^2 + \left(1+\phi'\right)^2\cos^2\theta + \left(3\sin^2\phi\cos^2\theta - 1\right)\right], \qquad (12)$$

$$u_3 = \rho\left[3C\left(C - 4\left(\theta'^2 + \phi'\cos^2\theta\left(1+\phi'\right)\right)\right)\right], \qquad (13)$$

as the nonlinear tension control law, so that

$$V' = -K_2\rho'^2, \quad K_2 > 0. \qquad (14)$$

One can see from equation (14) that $V' \leq 0$ with $V' = 0$ if and only if $\rho' = 0$. Thus, $V'$ is negative semi-definite and the controlled system is stable about the desired final state $\rho = \rho_f$, $\phi = \frac{\pi}{2}$, $\theta = 0$, $\rho' = \phi' = \theta' = 0$[42]. Finally, based on the stability condition of the system, the gains selected in this study were $K_1 = 2$, $K_2 = 6$ during deployment, and $K_1 = 1$, $K_2 = 6$ during retrieval.

## Acknowledgements

The authors thank the financial support of the CAPES (Coordination for the Improvement of Higher Education Personnel,Brazil), Grant 88887.478205 /2020-00, and Fapesp (São Paulo Research Foundation, Brazil), Grant 2016/24561-0.